\documentclass{elsart}

\usepackage{amssymb}

\usepackage[centertags]{amsmath}
\usepackage{amsfonts}
\usepackage{newlfont}

\numberwithin{equation}{section}

\newcommand{\opunit}{\text{1}\kern-0.22em\text{l}}

\begin{document}

\begin{frontmatter}



\title{Quantum entropy production as a measure of irreversibility}


\author{I. Callens},
\author{ W. De Roeck\thanksref{fwo}},
\author{ T. Jacobs},
\author{ C. Maes\corauthref{email}},
\author{K. Neto\v cn\'y}
\thanks[fwo]{Aspirant FWO, U.Antwerpen}
\corauth[email]{Christian.Maes@fys.kuleuven.ac.be}
\address{Instituut voor Theoretische Fysica\\ K.U.Leuven, Belgium.
}

\begin{abstract}
 We consider
conservative quantum evolutions possibly interrupted by
macroscopic measurements. When started in a nonequilibrium state,
the resulting path-space measure is not time-reversal invariant
and the weight of time-reversal breaking equals the exponential of
the entropy production. The mean entropy production can then be
expressed via a relative entropy on the level of histories. This
gives a partial extension of the result for classical systems,
that the entropy production is given by the source term of
time-reversal breaking in the path-space measure.
\end{abstract}

\begin{keyword}
quantum systems, entropy production, time-reversal
\PACS 05.30.Ch \sep 05.70.Ln
\end{keyword}
\end{frontmatter}

\section{Quantum entropy}
\label{quen}

In bridging the gap between classical mechanics and thermal
phenomena, the founding fathers of statistical mechanics realized
that a statistical characterization of macrostates in terms of
microstates is essential for understanding the typical time
evolution of thermodynamic systems. But statistical reasoning
requires proper counting, starting from a set of {\it a priori}
equivalent microstates. That is different in quantum statistics
from what is done in classical statistics.  Nevertheless the same
concepts can be put in place.  In this paper, we show for the
quantum case what has been proven useful in the classical case:
that entropy production measures the breaking of time-reversal
invariance.  More exactly, we show that the logarithmic ratio of
the probability of a trajectory and its time-reversal coincides
with the
physical entropy production.\\
We continue with an introduction about entropy, both classical and
quantum, to specify what we mean by entropy production. The main
message is contained in Sections 2 and 3 where the relation
between entropy production and time-reversal is demonstrated.
Section 2 deals with a unitary evolution (one initial preparation
and one final measurement) and can be considered as a special case
of Section 3 where the unitary evolution is further interrupted
with measurements of the macrostate. The paper is based on
elementary mathematical manipulations and the emphasis is on the
conceptual framework.

\subsection{Classical set-up}

Suppose  a classical closed and isolated system of $N$ particles.
A complete description consists in specifying its microstate, that
is a point $x$ in its phase space,
$x=(q_1,\ldots,q_N,p_1,\ldots,p_N)$ of positions and momenta.  The
dynamics is Hamiltonian and energy is conserved. Each microstate
$x$ determines a macrostate $M(x)$ corresponding to a much coarser
division of phase space.  To be specific, we consider the surface
$\Gamma$ of constant energy and we suppose that the macrostates
induce a finite partition $\hat{\Gamma}$ of $\Gamma$.  The
different macroscopic values thus specify a (finite) number of
regions in phase space.  We emphasize that this reduced
description is in terms of macroscopic variables (in contrast with
the partitions as used e.g in constructions of dynamical
entropies), such as position or velocity profiles and
we denote by $|M|$ the corresponding phase space volume. \\
  The
Boltzmann entropy (also called Boltzmann-Planck-Einstein or
microcanonical or configurational entropy) is, up to
multiplicative and additive constants, defined by
 \begin{equation}\label{BE}
 S_B(x) \equiv \log
|M(x)|
 \end{equation}
When the microstate belongs to the region of phase space of
unconstrained equilibrium, then, at least in the thermodynamic
limit,  $S_B(x)$ equals the thermodynamic entropy defined
operationally by Clausius.  But also out of equilibrium, say when
a constraint is lifted and the system is free to evolve, the
Boltzmann entropy remains relevant as it will typically increase
towards equilibrium.  It gives in fact the microscopic foundation
of the second law of thermodynamics.\\  There is another entropy
that generalizes \eqref{BE}. Suppose we only specify a
distribution of macrovalues. This means that we do not know the
exact macrostate, perhaps not even initially. The statistics of
macroscopic values is then given by a probability distribution
$\hat\nu(M), M\in \hat\Gamma$; e.g. our best guess about the
position and velocity profile.  In the absence of any further
information, there is a natural probability density, written
$\rho_{\hat\nu}$, on $\Gamma$ which gives equal probability to
each microstate compatible with a macrostate $M$:
 \[
\rho_{\hat\nu}(x) = \frac{\hat\nu(M(x))}{|M(x)|}
\]
The functional
\[
S_G(\hat\nu) = -\int_\Gamma dx \,\rho_{\hat\nu}(x) \ln
\rho_{\hat\nu}(x)
\]
is appropriately called Gibbs entropy as it is, since Gibbs, the
usual statistical mechanical entropy for constrained
equilibrium\footnote{Often, even without a specific physical
context or meaning, $-\int_\Gamma dx\,\mu(x) \log \mu(x)$ is
called the Gibbs entropy of $\mu$. There is however another name
for that functional: the Shannon entropy. A way out would be to
call $S_G(\hat\nu)$ the Boltzmann-Gibbs-Shannon entropy as in
\cite{W} but that sounds overdone.}. Put differently, by Gibbs'
variational principle,
\begin{equation}\label{gibbs}
S_G(\hat{\nu}) = \sup_{p(\mu)=\hat\nu} - \int_\Gamma dx \,\mu(x)
\ln \mu(x)
\end{equation}
where the supremum is taken over all phase space densities $\mu$
with the same macro-statistics as $\hat\nu$, i.e., the projection
$p(\mu)$ of $\mu$ on $\hat\Gamma$ coincides with $\hat{\nu}$:
 \[
  \mu(M) = \int_{\Gamma} \,
dx \, \mu(x) \,\delta(M(x)-M) = \hat\nu(M)
\]
and the supremum is reached for $\mu=\rho_{\hat\nu}$.
Equivalently, by a simple calculation, the Gibbs entropy equals
\begin{equation}\label{GE}
S_G(\hat\nu) \equiv  \sum_M \hat\nu(M) \log |M| - \sum_M
\hat\nu(M) \log \hat\nu(M)
\end{equation}
Obviously, if $\hat\nu$ concentrates on just one macrostate $M$,
$\hat\nu(M') = \delta_{M,M'}$, then $S_G(\hat\nu) = \log |M|$
which is the Boltzmann entropy for that macrostate. More
generally, the first term in \eqref{GE} is the expected or
estimated entropy, being unsure as we are about the exact
macrostate.  The second term in \eqref{GE}  is non-negative and is
most often negligible, certainly upon dividing by the number of
particles.\\
 When we start the system from a microstate that is
randomly drawn (according to the Liouville measure) from $M$ with
probability $\hat{\nu}(M)$ and we record the statistics
$\hat{\nu}_t$ on the macrostates after time $t$, then, always,
\begin{equation}\label{not2}
S_G(\hat\nu_t) \geq S_G(\hat\nu)
 \end{equation}
  This is not the second law
(e.g. it does not imply that $S_G(\hat\nu_t)$ is non-decreasing in
$t$, see \cite{mn} for more discussion). To bring it closer to the
second law, one needs to explain under what conditions
$S_G(\hat\nu_t)$ approximates the Boltzmann entropy $S_B(x_t)$
with $x_t$ the time-evolved phase point. Looking back at the first
term in \eqref{GE}, that implies understanding how and when the
empirical distribution for the macrovariables becomes peaked, as
the number of particles increases, at a fixed macroscopic
trajectory as described e.g. by
hydrodynamic equations of irreversible thermodynamics.\\
In \cite{mn} we have discussed at length how these entropies
relate to time-reversal.  The main result there was that both for
closed systems and for open systems, be it  in the transient
regime or in the steady state regime, the entropy production
equals the source term of time-reversal breaking in the action
functional for the distribution of the histories (on some
thermodynamic scale) of the system. This representation  gives the
entropy production as a function of the trajectories and it allows
simple manipulations for taking the average (to prove that it is
positive) and for studying its fluctuations (to understand
symmetries under time-reversal). In the present paper we begin
explaining a similar reasoning for finite quantum systems.

\subsection{Quantum set-up}

We take the simplest mathematical set-up for describing a quantum
system within conventional quantum mechanics.  We denote by
$\cal{H}$ the Hilbert space of the system, assumed finite
dimensional.  The system is described in terms of a wavefunction
$\psi$, or perhaps better, a normalized vector in $\cal{H}$. The
variables defining the macrostate are not different from that in
the classical situation.  For example, they specify (again to some
appropriate accuracy) the particle number and the momentum
profiles by associating numbers to a family of macroscopically
small but microscopically large regions of the volume occupied by
the system. Therefore, the macrostate is given in terms of the
values of a set of macroscopic observables represented by
commuting operators.  The macroscopic partition in the classical
case above is replaced by the orthogonal decomposition
\begin{equation}\label{decomp}
\cal{H} = \bigoplus \cal{H}_\alpha
\end{equation}
into linear subspaces.  The macrovariables are represented by the
projections $P_\alpha$ on the respective $\cal{H}_\alpha, P_\alpha
P_\beta = \delta_{\alpha,\beta} P_\alpha, \sum_\alpha P_\alpha = $
id. We write $d_\alpha$ for the dimension of $\cal{H}_\alpha$; it
is the analogue of the phase space volume $|M|$ in the classical
case.  We can think of the macrostates as labeled by the running
index $\alpha$.  For a given macrostate $\alpha$ represented by
the projection $P_\alpha$ above, we stretch the notation of
\eqref{BE} and write its Boltzmann entropy as
\begin{equation}\label{QBE}
\widehat{S_B}(\alpha) \equiv \log d_\alpha \equiv \log d(P_\alpha)
\end{equation}

So far, the above does not depart essentially from the classical
set-up except for the important difference that \eqref{BE} is
defined on microstates and \eqref{QBE} is defined on macrostates.
At this moment the treatment starts to differ.
 Following
von Neumann, page 411 in his well-known book \cite{vN}, as
discussed e.g. in \cite{Leb}, the formula \eqref{BE} is changed to
\begin{equation}\label{NE}
S_N(\psi) = \sum_\alpha (\psi,P_\alpha \psi) \log d_\alpha -
\sum_\alpha (\psi,P_\alpha \psi) \log (\psi,P_\alpha \psi)
\end{equation}
where $(\psi,P_\alpha \psi)$ is the probability of finding the
system described via the wavefunction $\psi$ in the macrostate
$\alpha$ (values corresponding to $\cal{H}_\alpha$).  Observe that
\eqref{NE} very much resembles \eqref{GE}, eventhough \eqref{NE}
is the quantum analogue of \eqref{BE}, both defined as they are
for the microstate. The reason is of course quantum mechanics
itself:  the wavefunction can still correspond to different
(mutually exclusive) macrostates.  If not initially, still later
in time, superpositions in terms of wavefunctions of different
macrostates are possible. This in contrast with the classical
set-up where a unique macrostate is associated to each microstate.
Just as the second term in the classical \eqref{GE} arises from
additional uncertainty as to the exact macrostate,  here, even
when the microstate is given (in terms of $\psi$), we still have a
distribution on the macrostates \footnote{One of us, C.M., likes
to emphasize that all this does not exclude that the actual system
is all the time in exactly one of the macrostates
--- but the wavefunction does not provide a complete description
of the system.}.

Instead of taking a pure state in \eqref{NE}, we may want to
consider the system in a mixed state and described by a density
matrix. In fact, as we are interested in the thermodynamic
time-evolution, we construct a special class of density matrices
by specifying the macroscopic statistics just as we did for
\eqref{GE}.  We thus start from a probability distribution
$\hat\mu$ on the possible macrovalues and we randomize within.
More precisely, we consider density matrices $\rho(\hat\mu)$
defined by
\begin{equation}\label{mus}
\rho(\hat\mu) \equiv \sum_\alpha \frac{\hat\mu(\alpha)}{d_\alpha}
P_\alpha
\end{equation}
for given $\hat{\mu}(\alpha) \geq 0, \sum_\alpha\hat{\mu}(\alpha)
= 1$. Then,
\begin{equation}\label{probs}
\hat\mu(\alpha) = \mbox{Tr}[ P_\alpha \,\rho(\hat\mu) ]
\end{equation}
 is the probability to find the
system in macrostate $\alpha$.  For the choice $\hat\mu(\alpha) =
d_\alpha/d$ where, for normalization, $d\equiv \sum_\alpha
d_\alpha$ is the dimension of $\cal{H}$,
$\rho(\hat\mu) = $ id$/d$.\\
In the other direction, given a general density matrix $\rho$, we
can take its projection $p(\rho)$ on the macroscopic states:
\begin{equation}\label{proj}
p(\rho)(\alpha) \equiv \mbox{Tr}[ P_\alpha \,\rho ]
\end{equation}
and of course, $p(\rho(\hat\mu))=\hat\mu$.

We now apply the same principle as in \eqref{gibbs} to obtain the
(Gibbs - von Neumann) quantum entropy
\begin{equation}\label{Qgibbs}
S(\hat\mu) \equiv \sup_{p(\rho)=\hat\mu} - \mbox{ Tr} [\rho \,\log
\rho]
\end{equation}
with solution (reached at $\rho= \rho(\hat\mu)$)
\begin{equation}\label{rQGE}
S(\hat\mu) = \sum_\alpha \hat\mu(\alpha)  \log d_\alpha -
\sum_\alpha \hat\mu(\alpha) \log \hat\mu(\alpha)
\end{equation}
which looks exactly like \eqref{GE}.  Again, $S(\hat\mu)$ equals
$\widehat{S_B}(\alpha)$ when $\hat\mu$ concentrates on the
macrostate $\alpha$ and, up to some irrelevant constants,
$S(\hat\mu)$ equals the relative entropy of $\rho(\hat\mu)$ with
respect to the equilibrium density matrix  id$/d$.
\\ More
generally,  we call
\begin{equation}\label{QGE}
S(\rho) \equiv S(p(\rho)) = \sum_\alpha \mbox{ Tr} [P_\alpha
\,\rho] \log d_\alpha - \sum_\alpha  \mbox{ Tr} [P_\alpha\,
\rho]\,\log \mbox{Tr} [P_\alpha\, \rho]
\end{equation}
the macroscopic quantum entropy of the (mixed) state $\rho$,
reducing to \eqref{rQGE} for the choice \eqref{mus} and to
\eqref{NE} for a pure state.

\section{Entropy production for a unitary evolution}

The goal of this section is to show that the total change of
entropy under a quantum conservative evolution, that is the
entropy production, can be given in terms of a logarithmic ratio
of probabilities.

 We suppose a quantum dynamics generated via a
real Hamiltonian $H$.  This means that the Hamiltonian is
time-reversal invariant, $H\pi =\pi H$ where $\pi$ is an
anti-linear involution (kinematical time-reversal) which would be
just complex conjugation for wavefunctions.  The unitary evolution
$U(t) \equiv \exp[-itH]$ where $t$ is time (and Planck's constant
is set equal to one) then satisfies
\[
 \pi U(t) \pi = U(t)^\star = U(t)^{-1}
\]
an expression of dynamic reversibility.   We also assume that the
macrostates are mapped into each other via the involution $\pi$,
i.e.,
 \begin{equation}\label{consa}
 \pi
P_\alpha \pi = P_{\alpha'} \equiv P_{\pi\alpha}
\end{equation}
 for some $\alpha'$, for each $\alpha$
and we write $\pi\alpha =\alpha'$.

Let the system be initially prepared with macro-statistics
$\rho(\hat\mu)$, see \eqref{mus}.  We now leave the system alone,
undergoing its quantum evolution.  The probability to see the
system at the initial time in macrostate $\alpha_0$ and at time
$t$ in macrostate $\alpha_t$ is given by
\begin{equation}\label{traj1}
\mbox{Prob}_{\hat\mu}[\alpha_0,\alpha_t] = \mbox{Tr
}[U(t)P_{\alpha_0}\rho(\hat\mu)P_{\alpha_0} U(t)^\star
P_{\alpha_t}]
\end{equation}
Its marginal at time $t$ is
\begin{equation}\label{n=1}
\hat\mu_t(\alpha) \equiv \sum_{\alpha_0}
\mbox{Prob}_{\hat\mu}[\alpha_0, \alpha_t=\alpha] = \mbox{Tr
}[U(t)\rho(\hat\mu) U(t)^\star P_\alpha]
\end{equation}

We consider the logarithmic ratio of probabilities
\begin{equation}\label{logtraj1}
R_{\hat\mu}(\alpha_0,\alpha_t) \equiv
\log\frac{\mbox{Prob}_{\hat\mu}[\alpha_0,
\alpha_t]}{\mbox{Prob}_{\hat\mu_t\pi}[\pi\alpha_t, \pi\alpha_0]}
\end{equation}
where the denominator gives the probability of the time-reversed
order of measurement results starting from $\rho(\hat\mu_t\pi)$
with $\hat\mu_t\pi(\alpha) \equiv \hat\mu_t(\pi\alpha)$.

We make two observations.  First, when $\hat\mu(\alpha) =
d_\alpha/d$, then also at later times $\hat\mu_t(\alpha) =
d_\alpha/d$ and moreover
\begin{equation}\label{star}
\mbox{Prob}_{\hat\mu}[\alpha_0,\alpha_t] =
\mbox{Prob}_{\hat\mu_t}[\pi\alpha_t,\pi\alpha_0]
\end{equation}
expressing the time-reversal invariance when started from the
time-invariant macro-statistics that assigns to macrostates
probabilities proportional to the exponential of their Boltzmann
entropy, see \eqref{QBE}.\\
Secondly, for other choices of the initial macro-statistics, the
$R_{\hat\mu}$ of \eqref{logtraj1} reproduces the change of
entropy:
\begin{equation}\label{ent1}
R_{\hat\mu}(\alpha_0,\alpha_t) = \log d_{\alpha_t} - \log
d_{\alpha_0} - \log \hat\mu_t(\alpha_t) + \log \hat\mu(\alpha_0)
\end{equation}
where we recognize the change of Boltzmann entropies \eqref{QBE}
in the first two terms. Upon taking the expectation
\begin{equation}\label{ment1}
\sum_{\alpha_0,\alpha_t}\mbox{Prob}_{\hat\mu}[\alpha_0, \alpha_t]
\, R_{\hat\mu}(\alpha_0,\alpha_t) = S(\hat\mu_t) - S(\hat\mu)
\end{equation}
we recover the change in entropy \eqref{Qgibbs}.  Note that
$S(\hat\mu_t) = S(\rho_t), S(\hat\mu)=S(\rho)$ defined in
\eqref{QGE} for $\rho=\rho(\hat\mu)$, $\rho_t= U(t)\rho
U(t)^\star$ so that the right-hand side of \eqref{ment1} is really
the change of macroscopic quantum entropy under the unitary
evolution.  The left-hand side of \eqref{ment1} is a relative
entropy, hence is non-negative.

\section{Perturbations by measurement}\label{psm}

In contrast with the classical evolution, the  unitary evolution
can be non-trivially interrupted by measurements that reduce the
state. We follow here the theory of von Neumann (or, convential
quantum mechanics) that adds a dynamic interpretation to the
projections \eqref{proj}.  We extend here the results of the
previous section to include interactions with a measurement
apparatus.\\
   Suppose that we start the system described by a
density matrix $\rho$ at time 0 and that we measure the
macroscopic value $\alpha$ at time $\tau>0$.  Then, the new
density matrix is
\[
\frac{P_\alpha U\,\rho\,U^\star P_\alpha}{\mbox{Tr}[P_\alpha
U\,\rho\,U^\star]}
 \]
 where $U \equiv U(\tau) = \exp[-i\tau H]$.
 This
reduction takes place with probability
\[
p(U\rho U^\star)(\alpha) = \mbox{Tr}[P_\alpha U\,\rho\,U^\star]
\]
We can generalize this to a sequence of measurements.

\subsection{Path-space measure}
We consider a sequence of times $0,\tau,2\tau,\ldots,n\tau=t$
(evenly spaced for convenience). A macroscopic trajectory $\omega$
assigns to each of these times a value for the macroscopic
observables. In short, $\omega=(\omega_0,\ldots,\omega_n)$ with
each $\omega_i$ being equal to some projection $P_\alpha$ on the
linear subspace $\cal{H}_\alpha$ of \eqref{decomp}. We consider
the
operator
\begin{equation}\label{gobj}
G_\omega \equiv \omega_n U \omega_{n-1}\ldots\omega_1 U \omega_0
\end{equation}
and, for a given density matrix $\rho$ on $\cal{H}$, the matrix
\begin{equation}\label{deco}
\cal{D}_\rho(\omega,\omega') \equiv \mbox{ Tr}[G_\omega \,\rho\,
G_{\omega'}^\star]
\end{equation}
This resembles the correlation matrix that is used in the
definition of quantum dynamical entropy, see p. 189 in \cite{AF},
but again, here our partitioning corresponds to macrostates.\\
 It is easy
to
 verify that $\cal{D}_\rho$ is non-negative and has trace one.
We call it the path-space density matrix.\\
$\cal{D}_\rho$ depends on the initial density matrix $\rho$ and it
also gives the probability to find the system after time $t=n\tau$
of (measurement-)free evolution in the macrostate $\alpha$:
\begin{equation}\label{endrho}
\sum_{\omega,\omega': \omega_n=\omega_n'=P_\alpha}
\cal{D}_\rho(\omega,\omega') = \mbox{ Tr}[P_\alpha \; \rho_t]
\end{equation}
where $\rho_t \equiv U^n \rho U^{-n} $.\\  More importantly, the
probability to measure the trajectory $\omega$, i.e., to see the
system at the initial time in the macrostate represented by
$\omega_0$, at time $\tau$ in macrostate $\omega_1$, and so on
 till time $n\tau=t$, is given by the diagonal element
\begin{equation}\label{macroprob}
 \mbox{Prob}_\rho[\omega] = \cal{D}_\rho(\omega,\omega)
\end{equation}
Its marginal at time $t=n\tau$,
 \begin{equation}\label{marg}
\hat\mu_t(\alpha) \equiv \sum_{\omega: \omega_n=P_\alpha}
\mbox{Prob}_\rho[\omega]
\end{equation}
is the probability of macrostate $\alpha$ when the unitary
evolution was interrupted by $n$ measurements. It equals
\eqref{n=1} in case $n=1$.

\subsection{Time-reversal}
 On trajectories, the
time-reversal transformation $\Theta$ is
\[
(\Theta\omega)_m \equiv \pi \omega_{n-m} \pi,\quad m=0,1,\ldots,n
\]
Since the microscopic dynamics is time-reversal invariant, we
immediately deduce that
\[
G_{\Theta\omega} = \pi G_\omega^\star\pi
\]
and it is easy to verify that for $\rho = $ id$/d$ in \eqref{deco}
\begin{equation}\label{timerevid}
\cal{D}(\Theta\omega',\Theta\omega) = \cal{D}(\omega,\omega')
\end{equation}
In particular, for \eqref{macroprob} with $\rho = $ id$/d$,
Prob$[\omega]=$ Prob$[\Theta\omega]$. This identity has two
related interpretations. First, it is the expression of (quantum)
detailed balance or microscopic reversibility for the trajectories
sampled from the time-invariant density matrix id$/d$
corresponding to equilibrium. It is the generalization of
\eqref{star}. Secondly, it generalizes the observation of
\cite{ABL} that measurements do not introduce a time-asymmetric
element (or, the reduction of the wave packet does not lead to
irreversibility).\\
Yet, in general, the system could start in a nonequilibrium state
and evolve towards equilibrium. This involves a change of entropy;
that is, for closed and thermally isolated systems, its total entropy production.\\
 The
time-reversal of the density matrix $\cal{D}_\rho$, written
$\cal{D}_\rho\Theta$, is defined as
\begin{equation}\label{timerevgen}
\cal{D}_\rho\Theta(\omega,\omega') \equiv
\cal{D}_\rho(\Theta\omega',\Theta\omega)
\end{equation}
so that by \eqref{timerevid}, $\cal{D}\Theta = \cal{D}$. This last
equality is broken for a general $\cal{D}_{\rho}$ and it seems
natural to estimate this breaking via the relative entropy
\begin{equation}\label{relen}
S(\cal{D}_{\rho_1}|\cal{D}_{\rho_2}\Theta) \equiv
 \mbox{Tr}[\cal{D}_{\rho_1} (\log\cal{D}_{\rho_1} -\log
 \cal{D}_{\rho_2}\Theta)]
 \end{equation}
In view of the classical results of \cite{m,mn}, it is to be
expected that this relative entropy is related to the entropy
production for the appropriate choices of $\rho_1$ (as initial
state) and of $\rho_2$ (as time-reversal of the final state).

\subsection{Entropy production}\label{enter}

We start again from $\hat\mu$ as initial probability distribution
(i.e., $\rho(\hat\mu)$ is the initial density matrix) and the
final probability distribution $\hat\mu_t$ is defined in
\eqref{marg}.
The main result is now readily obtained starting with the analogue
of \eqref{logtraj1}: the logarithmic ratio of probabilities
\eqref{macroprob} of a macroscopic trajectory, one started from
$\hat\mu$ and the other started from $\hat\mu_t\pi$, is
abbreviated as
\begin{equation}\label{mainnot}
R_{\hat\mu}(\omega) \equiv \log
\frac{\mbox{Prob}_{\hat\mu}[\omega]}{\mbox{Prob}_{\hat\mu_t\pi}[\Theta\omega]}
\end{equation}
This object will not always be well-defined for all $\omega$.  We
can suppose however that the $\hat\mu(\alpha) \neq 0 \neq
\hat\mu_t(\alpha)$ so that a little calculation of \eqref{mainnot}
yields
\begin{equation}\label{main}
R_{\hat\mu}(\omega)  = \log d(\omega_n) - \log d(\omega_0) - \log
\hat\mu_t(\omega_n) + \log \hat\mu(\omega_0)
\end{equation}
In the notation of \eqref{QBE}, the first difference in the
right-hand side of \eqref{main} equals the change of Boltzmann
entropy $\widehat{S_B}(\alpha_t) - \widehat{S_B}(\alpha_0)$ for a
trajectory that
 starts in
macrostate $\alpha_0$ and ends in macrostate $\alpha_t$.  If the
system is initially prepared in essentially one macrostate,
$\hat\mu(\alpha_0)\simeq 1$, and if the evolution on the level of
macrostates is quasi-autonomous till time $t$ in the sense that
$\hat\mu_t(\alpha_t) \simeq 1$, then only that change in Boltzmann
entropies
survives. \\
Taking the expectation of \eqref{mainnot}, we get, similar as in
\eqref{ment1},
\begin{equation}\label{expmain1}
\sum_\omega \mbox{Prob}_{\hat\mu}[\omega]\,R_{\hat\mu}(\omega) =
S(\hat\mu_t) - S(\hat\mu) \geq 0
\end{equation}
equal to the change of quantum entropy
\eqref{Qgibbs}--\eqref{rQGE}. The non-negativity follows from
Jensen's inequality applied to the left-hand side of
\eqref{expmain1}.\\
 One can easily check that this expectation
(the left-hand side of \eqref{expmain1}) and hence, the change of
quantum entropy (the right-hand side of \eqref{expmain1}), also
coincides with the relative entropy on the level of density
matrices for trajectories
\[
S(\cal{D}_{\hat\mu}|\cal{D}_{\hat\mu_t\pi}\Theta) = \mbox{Tr
}[\cal{D}_{\hat\mu}\,(\log \cal{D}_{\hat\mu} - \log
\cal{D}_{\hat\mu_t\pi}\Theta)]
\]
as announced in \eqref{relen}.

\section{Conclusions and additional remarks}

Entropy production, also for the quantum situation, has been under
intense investigation in recent years.  A list of related
references, even restricted to the last 5 years would soon fill an
extra couple of pages.  In the classical case, there have been
essentially two types of approaches, dynamical versus statistical.
In the dynamical approach, entropy production is connected with
the fractal or singular nature of a class of natural stationary
measures.  A key-notion is the phase space contraction rate and
with some extrapolations, nonequilibrium statistical mechanics
becomes a branch of the theory of smooth dynamical systems. It was
partially motivated by numerical work and it has given wonderful
insights, guiding towards a study of fluctuations of the entropy
production, possibly generalizing close to equilibrium relations.
The statistical approach emphasizes the sharp contrast between the
microscopic and macroscopic scales and stochasticity enters the
dynamical evolutions because of the reduced description. On
space-time the histories have a distribution that share the
essential properties of Gibbs measures. The entropy production can
then be identified with the source term of time-reversal breaking
in the action (or space-time Lagrangian) of the space-time
distribution; the Gibbs formalism reproduces and extends the
results for the fluctuations of the entropy production obtained in
the purely dynamical approach.  The projections on equal time
surfaces reproduce the time-evolved measure. In the quantum case,
the modern dynamical approach does not seem to have a natural
generalization. Instead, the old machinery of operator-algebra and
spectral theory, directly applied on infinite volume systems, has
been providing encouraging results for relaxation to equilibrium
and the positivity of entropy production in driven quantum
systems. Again however, the conceptual framework and the essential
distinction between microstates and macro-statistics is, at least
to our taste, most often blurred in non-intuitive dynamical
assumptions and in the heavy mathematical formalism. The present
paper offers an alternative and starts the statistical approach
for quantum systems.  The first step indeed is to be quite clear
as to the relation between time-reversal and entropy production.
As in the classical case, we consider this important for
constructing nonequilibrium statistical mechanics. Specifically,
we hope that the characterizations of \eqref{ent1} and
\eqref{main} will prove useful for deriving general fluctuation
and response
identities going beyond close to equilibrium.\\
We add some general remarks.

1. For reversible systems the entropy production is a state
function. It does not depend on the actual path but can be written
as a difference of the same quantity evaluated at the initial and
the final time, see \eqref{ent1} or \eqref{main}.  This can be
different for driven or irreversible systems.  A similar treatment
for such systems in the classical regime can be found in
\cite{mn}. One of the main consequences of the identities that
correspond there to \eqref{ent1} or \eqref{main} is that certain
symmetries in the fluctuations of the entropy production rate are
easily established. The quantum case is in preparation.

2. Nothing of the previous mathematical identities depends on the
assumption that the system is large, of macroscopic size,
containing a huge number of particles.  Of course, to meaningfully
discuss macroscopic variables or macrostates and their
trajectories one has in mind a clear separation between micro and
macro scales but the results of Sections 2 and 3 are true without.
Yet again, their interpretation as thermodynamic entropy
production and the relation with the second law can only be made
when dealing with macroscopic systems. See  also Appendix A in
\cite{mn} for these considerations in the classical case.

3.  The use of the projections $P_\alpha$ in the construction of
the path-space density matrix of Section \ref{psm} refers to the
so called von Neumann measurements.  One can imagine more {\it
fuzzy} measurements (and more rounded-off macrostates) and a
corresponding decomposition of unity as
\[
\sum_\alpha X_\alpha^\star X_\alpha = \mbox{ id}
\]
The treatment above was restricted to the case $X_\alpha =
P_\alpha U$. This extension is very much related to the dynamics
of open systems, see e.g. \cite{AF}. The involvement of
measurements interrupting the unitary evolution already points to
the interaction with the outside world. The question of isolation,
even as an idealization, of a quantum system is much more subtle
than for a classical system.

\end{document}